# Heterogeneous teleportation with laser and quantum light sources

R. M. Stevenson[1]*, J. Nilsson[1], A. J. Bennett[1], J. Skiba-Szymanska[1], I. Farrer[2], D. A. Ritchie[2], A. J. Shields[1]*

[1] Toshiba Research Europe Limited, 208 Science Park, Milton Road, Cambridge, CB4 0GZ, United Kingdom.

[2] Cavendish Laboratory, University of Cambridge, J. J. Thompson Avenue, Cambridge, CB3 0HE, United Kingdom.

*Correspondence to: mark.stevenson@crl.toshiba.co.uk, andrew.shields@crl.toshiba.co.uk

**Abstract**

Quantum information technology is set to transform critical network security using quantum cryptography, and complex scientific and engineering simulations with quantum computing. Quantum computer nodes may be based on a variety of systems, such as linear optics[1,2], ions [3,4] or solid state architectures such as NV-centers in diamond [5,6], semiconductor quantum dots [7,8] or spins in silicon [9]. Interfacing any of these platforms with photonic qubits in secure quantum networks will require quantum teleportation protocols to transfer the information, and matter-light teleportation has for some of these systems been demonstrated [10,11]. However, although it is conceivable that the input photon originates from a dissimilar source to that supplying the entangled resources, every demonstration so far of teleportation using linear optics use the same [12,13] or identical [14,15] sources for the input and entangled photons, often accompanied by a fourth heralding photon[14]. Here we show that photons from fundamentally different sources can be used in the optical quantum teleportation protocol. Input photons are generated by a laser, and teleported using polarisation-entangled photon pairs electrically generated by an entangled-light-emitting diode (ELED)[16]. The sources have bandwidth differing by a factor $10^3$, different photon statistics and need not be precisely degenerate- but we still observe a teleportation fidelity of 0.77, beating the quantum limit by 10 standard deviations. This is a significant leap towards practical applications, such as extending the range of existing QKD systems using quantum relays [17] and repeaters [18], which usually use weak coherent laser pulses for quantum information transport. The use of an ELED offers practical advantages of electrical control, and as we show erases the multi-photon character of the laser input field, thus eliminating errors if used in a quantum optics circuit.

**Main Text**

Linear optics quantum teleportation requires Hong-Ou-Mandel (HOM) type interference [19,20], usually realised using a 50:50 beamsplitter, between the input qubit and one ancilla photon from an entangled photon pair [12,21]. Here, we interfere our photons on a 95:5 unbalanced beamsplitter to perform quantum teleportation of the polarisation state carried by input photons from a CW laser. The large coupling imbalance allows us to make efficient use of the photons produced by the quantum dot emitter.

Before proceeding to the quantum teleportation experiments, we verify the two-photon interference of our dissimilar light sources on the unbalanced beamsplitter using the setup illustrated schematically in Fig. 1.a. The setup is implemented using single mode fibre components and single-photon counting detectors. XX photons from the ELED are fed into input port $a_1$ of the unbalanced 95:5 beamsplitter, which couples with 95% efficiency to output mode $a_3$, and the CW laser is fed to input port $a_2$ with coupling 5% to port $a_3$. Using a balanced beamsplitter (50:50) at port $a_3$ we measure second-order correlations using detectors D1 and D2 for co-polarised (interfering) and cross-polarised (non-interfering) inputs.

The measured second-order correlation functions $g_{\parallel}^{(2)}(\tau_1)$ for interfering photons are shown in Fig. 1.b for increasing detuning $\Delta E$. For zero detuning (bottom correlation) we also show the second-order correlations $g_{\perp}^{(2)}(\tau_1)$ for non-interfering, orthogonally polarised photons, which shows a clear dip originating from the partially sub-Poissonian photon stream, in contrast to the clear peak in the co-polarised correlation. We note that this peak originates from our direct observation of the "bunching" behaviour due to bosonic coalescence in $a_3$, in contrast to previous experiments that usually observe two-photon interference as an *absence* of coincidences in opposite output ports $a_3$, $a_4$ [19,20,22].

As we increase the detuning quantum beats with increasing frequency appear in the correlations, and to the best of our knowledge this is the first observation of beats of this kind for a quantum dot emitter. In Fig. 1.b one can see that the energy detuning effectively narrows the central peak in $g_{\parallel}^{(2)}(\tau_1)$, which leads to a reduced magnitude of the observed bunching as the beat period approaches the detector time resolution (80 ps for the pair D1-D2). Fig. 1.c summarises this effect in terms of observed and simulated peak interference visibility $V = \left(g_{\parallel}^{(2)}(0) - g_{\perp}^{(2)}(0)\right)/g_{\perp}^{(2)}(0)$. It is notable that the interference is surprisingly robust, with appreciable visibility still after 15 μeV, and beats still visible at 40 μeV, several times larger than the XX linewidth of $2\hbar/\tau_c \sim 8$ μeV. This indicates a robustness of the post-selective protocol implemented here, owing to the quantum eraser effect imposed by the post-selection.

To implement quantum teleportation we use the setup shown schematically in Fig. 2.a; on the input the ELED is no longer polarised, and the exciton photon (X) is now sent to a receiving node (Bob) equipped with a fibre-based polarising beamsplitter (PBS) and single-photon counting detectors. On the output of the 95:5 splitter the 50:50 beamsplitter is replaced with a PBS, calibrated to measure in the basis of the quantum dot exciton (X) eigenstates H-V. The 95:5 splitter, the PBS and the detectors D1 and D2 now constitute a Bell-state measurement (BSM) apparatus, which we will simply refer to as Alice from now on. A coincident detection by Alice ($\tau_1 = 0$) marks a successful BSM which projects the XX and laser photons at the input of the 95:5 splitter onto the Bell state $|\Psi^+\rangle = (|HV\rangle + |VH\rangle)/\sqrt{2}$, and signals the successful teleportation of the laser input polarisation state onto Bob's X photon (up to a trivial unitary transformation). The time $\tau_2$ of Bob's detection events is measured relative to the triggering of detector D1, and we record third-order correlation functions[23].

We simulate the performance of our teleporter for different laser intensities and detunings (see Methods for details) using a model that takes the finite photon interference, the quantum dot exciton fine-structure and Poissonian statistics of the laser into account. The results, shown in Fig. 2.b, suggest that in order to achieve less than 1% reduction of teleportation fidelity of a superposition input state such as $|D\rangle = (|H\rangle + |V\rangle)/\sqrt{2}$, the energy detuning needs to be less than ~10 µeV. We choose to perform the experiments at a quantum dot to laser intensity ratio (measured at D1 and D2) of $\eta/\alpha^2 = 2$ where fidelity is close to maximum.

We test the quantum teleportation protocol for six input laser polarisation states symmetrically distributed over the Poincare sphere in three polarisation bases; the rectilinear basis H/V coinciding with Alice's measurement basis and the quantum dot exciton eigenbasis, the diagonal basis spanned by $|D/A\rangle = 1/\sqrt{2}\,(|H\rangle \pm |V\rangle)$ and the circular basis $|R/L\rangle = 1/\sqrt{2}\,(|H\rangle \pm i|V\rangle)$. For each input state we measure the fidelity onto the expected output state by aligning Bob to the corresponding basis. For our choice of input states, the highest possible average output fidelity is 2/3 using the best possible classical teleporter [23]. For coincident detection by Alice and Bob ($\tau_1 = \tau_2 = 0$) we achieve 0.767 ± 0.012 as shown in Fig. 2.c, clearly beating the classical limit and proving that quantum teleportation is taking place. Also shown are cuts through the fidelity map at $\tau_1 = 0$ and $\tau_2 = 0$, together with a comparison with the model showing good agreement. At $\tau_1 = 0$ along Bob's time axis $\tau_2$ the peak width is limited by the XX-X polarisation correlations, and for $\tau_2 = 0$ along Alice's time axis $\tau_1$ the peak is limited by the XX coherence time $\tau_c$.

Fig. 2.d shows the simulated and experimentally measured fidelities for individual laser polarisation settings. The polar states on the Poincare sphere coinciding with Alice's measurement basis show the highest fidelity as expected ($f_{H\rightarrow V} = 0.835 \pm 0.026$, $f_{V\rightarrow H} = 0.861 \pm 0.024$), as these do not require successful interference in the Bell-state measurement apparatus, and could be done with classically correlated photon pairs. The four superposition states, which rely on both successful interference *and* entanglement, have similar output fidelities ($f_{D\rightarrow D} = 0.725 \pm 0.032$, $f_{A\rightarrow A} = 0.698 \pm 0.033$, $f_{R\rightarrow L} = 0.744 \pm 0.028$, $f_{L\rightarrow R} = 0.741 \pm 0.031$). The slightly higher fidelity in the circular basis

compared to the diagonal is consistent with polarisation correlations observed for this type of QD, which can be attributed to nuclear polarisation fluctuations in the quantum dot [16,24].

During the teleportation experiment we simultaneously perform a Hanbury Brown Twiss measurement to determine the second order correlation function of the X photons going down the optical fibre to bob. We find a characteristic dip with minimum $g^{(2)}_{X-X}(0) \sim 0.257 \pm 0.001$, which confirms that the setup erases the Poissonian statistical nature of the input laser field.

In the experiments described above Bob interrogates the output photon by measuring in the expected output polarisation basis only, which cannot reveal the full character of the output light. To explore this we perform single qubit tomography[25] of the output photon density matrix corresponding to input state R by measuring the output fidelity in the three bases H/V, D/A and R/L. For a perfect quantum teleportation one would expect input R to yield output L with unit fidelity and fidelity 0.5 in the other bases. At $\tau_1 = \tau_2 = 0$ we measure fidelities $f_{R \to L} = 0.713 \pm 0.031$, $f_{R \to D} = 0.646 \pm 0.034$ and $f_{R \to H} = 0.550 \pm 0.033$ from which we construct the real and imaginary parts of the output state density matrix shown in Fig. 3.a and 3.b. Of these states, the maximum fidelity is found for the expected output state L, and it is the same as in Fig. 2.d within the accuracy of the experiment, but the measurements also reveal a relatively strong D component which results in the non-zero off-diagonal imaginary components in Fig. 3.b.

The largest eigenvalue $\lambda_1$ of the density matrix tells us the maximum fidelity that can be measured, and the corresponding eigenvector $|v_1\rangle$ tells us the polarisation along which Bob should be aligned in order to measure this. Fig. 3.c shows $\lambda_1$ as a function of Bob's detection time $\tau_2$, with a peak value of 0.763±0.030. This exceeds the output fidelity to L, and confirms that indeed Bob's measurement was not optimally aligned to the actual output state. As $\tau_2$ increases, the X photon detected by Bob is no longer from the same radiative cascade as the XX photon detected by Alice, and $\lambda_1$ approaches 1/2 when the output becomes completely mixed. Experimentally, we can follow the evolution of the output state $|v_1\rangle$ up to $\tau_2 \sim 1$ ns after which the uncertainty becomes too large. Fig. 3.d depicts the overlap of this pure state with the desired state L and the orthogonal counterpart R, showing a clear evolution of the output from L towards R. The numerical model, in contrast to the experiment, is noise-free and a pure fraction (albeit still asymptotically vanishing for large $\tau_2$) can always be separated, and as shown in Fig. 3.d the predicted output state evolution is well-described by the fine-structure splitting (~2 μeV) of the X state, and agrees qualitatively well with the experimental observations.

Conclusion

To conclude, we have performed photonic quantum teleportation of input states encoded on photons from a coherent light source, teleporting them onto a stream of photons from a sub-Poissonian semiconductor emitter. With further improvements of the device design, such as placing the emitter in an optical nanocavity [26–28], the teleportation method presented here could find application in e.g. the realisation of quantum relays and repeaters for dissimilar light sources. The protocol used here, with a strongly unbalanced beamsplitter, could also provide a useful interface to remotely initialise quantum information processors using abundant laser-generated photons over long distances, and conserving more exotic sub-Poissonian light fields in the local quantum circuit. Other interesting applications could be to secure QKD networks, usually implemented using weak lasers, from Trojan horse attacks [29].

**Methods**

**Entangled-light-emitting diode (ELED).** The electrically excited entangled light source used in this study is comprised of self-assembled InAs/GaAs quantum dots in a p-i-n diode structure, sandwiched between distributed Bragg reflectors forming a weak planar optical cavity [30]. The device is operated at ~15 K in d.c. mode, passing a current through the junction and exciting the quantum dots at random times. The dot used here was verified to have a small exciton fine-structure splitting of 2.0 ± 0.2 µeV. With a Michelson interferometer we determine the coherence time $\tau_c$ of the biexciton photons (XX) to be 161 ± 4 ps at the driving current ~90 nA/µm$^2$ used.

**Fibre-based optical circuits.** The optical circuits schematically described in Figs. 1a and 2a were all implemented in single-mode fibre using unbalanced (95:5) and balanced (50:50) beamsplitters, polarising beamsplitters (PBS) and polarisation controllers. A tuneable spectral filter picks out the exciton (X) and biexciton (XX) photons without narrowing the transition linewidths. Polarisations in the fibre system were aligned to an external calibration laser beam coupled into the fibre system at the same point as the ELED emission.

For two-photon interference (Fig. 1a) and for the Bell measurement apparatus (Fig. 2a) superconducting single-photon counting detectors D1 and D2 were employed. In teleportation experiments Bob was in possession of avalanche photodiodes D3 and D4. All times were measured in relation to a triggering detection event on D1. The experimentally determined pair-wise detector resolutions were; D1-D2: 80 ps, D1-D3: 340 ps, D1-D4: 360 ps.

Port $a_4$ of the 95:5 beamsplitter (Figs. 1a and 2a) was used to monitor the laser and the biexciton spectral detuning. Through computer control of the laser cavity we were able to maintain a desired detuning with an estimated accuracy of ~5 µeV.

Polarising beamsplitters and electrical polarisation controllers at the sources (not shown) allow us to periodically alternate between measuring co- polarised (interfering) and cross-polarised (non-interfering) photons throughout the two-photon interference experiments. Similarly, in teleportation measurements the polarisation state of the input laser photons was periodically switched between orthogonal states (D-A, H-V and R-L).

**Error analysis.** The main source of uncertainties in all experiments is due to the Poissonian counting statistics. Quoted errors on teleportation fidelities also include an uncertainty in time calibration of the photon correlation equipment (less than 1% on individual fidelities). Errors in Fig. 3 (density matrix elements and eigenvalues) are estimated by propagating the counting statistics from the raw data.

**Modelling unbalanced two-photon interference.** The simulated two-photon interference visibilities for an unbalanced beamsplitter shown in Fig. 1c are based on a well-established wavepacket analysis [31,32]. By considering different cases that can lead to coincident detections at detectors D1 and D2 we can estimate the second-order correlation function for co-polarised laser and ELED [22]:

$$g^{(2)}_{\parallel}(\tau) = \frac{2\eta\alpha^2 \cdot \left(1 + e^{-|\tau|/\tau_c} \cdot \cos \Delta E \tau/\hbar\right) + \eta^2 g^{(2)}_{HBT}(\tau) + \alpha^4}{(\eta + \alpha^2)^2} \quad (1)$$

where $\eta$ is proportional to the XX photon intensity and $\alpha^2$ is proportional to the laser intensity measured at detectors D1 and D2, $\tau_c$ is the coherence time of the XX photons and $\Delta E$ is the XX to laser energy detuning. $g^{(2)}_{HBT}(\tau)$ is the XX transition second-order auto-correlation function, separately measured and analytically fitted. By convolving the above expression with the instrument response for detectors D1-D2 we arrive at the simulated interference visibility presented in Fig. 1c.

**Modelling heterogeneous quantum teleportation.** Detecting two photons on Alice's detectors D1 and D2, where D1 and D2 resolve orthogonal polarisations H and V in the same arm of the beamsplitter, effectively projects the detected photons onto the Bell state $|\psi^+_{12}\rangle$, compared to $|\psi^-_{12}\rangle$ in most quantum teleportation setups. Here 1 and 2 refer to the input ports of the beamsplitter, as labelled in Fig. 1a, and $|\psi^\pm_{12}\rangle = (|H_1V_2\rangle \pm |V_1H_2\rangle)/\sqrt{2}$, $|\phi^\pm_{12}\rangle = (|H_1H_2\rangle \pm |V_1V_2\rangle)/\sqrt{2}$ are the Bell states. Ideally, the exciton-biexciton pairs emitted by the quantum dot would be in the Bell state $|\phi^+_{XX,X}\rangle = (|H_{XX}H_X\rangle + |V_{XX}V_X\rangle)/\sqrt{2}$. With some algebra we find that for an arbitrary laser input polarisation $\alpha|H_2\rangle + \beta|V_2\rangle$ we should find the teleported state received by Bob to be $\alpha|V_X\rangle + \beta|H_X\rangle$. This means the following set of transformations imposed by the teleportation operation, which is consistent with the experimental results: H→V, V→H, D→D, A→A, R→L, L→R.

Using the same wavepacket analysis as for two-photon interference as a foundation, we can calculate the probability that Bob detects a certain polarisation state given a particular input laser polarisation [23]. As an example, the three-fold coincidence probability (Alice H,V, Bob D) for input state $|D_2\rangle = (|H_2\rangle + |V_2\rangle)/\sqrt{2}$ is

$$P_{HVD}(\tau_1,\tau_2) \propto e^{-2\gamma_x(\tau_2-\tau_1)} + e^{-2\gamma_x\tau_2} + 2e^{-|\tau_1|/\tau_c - \gamma_x(\tau_2-\tau_1) - \gamma_x\tau_2} \cdot \cos[s(\tau_2-\tau_1)/\hbar + \Delta E\tau_1/\hbar]$$

for detections by Bob at times later than both of Alice's detections ($\tau_2 > 0$ and $\tau_2 > \tau_1$). Here $\gamma_x$ is the lifetime of the X emitted after the XX photon. The effects of exciton fine-structure splitting (*s*) and laser-quantum dot detuning $\Delta E$ is apparent as oscillations in this expression, as well as the importance of the coherence properties of the XX photons ($\tau_c$).

In principle, the model above could let us calculate the fidelity of the detected photons for an idealised source, but $P_{HVD}(\tau_1, \tau_2)$ does not take the actual device driving conditions into account. As a full model of the quantum dot states and all associated transition rates is difficult to realise, we take a semi-empirical approach to predict the performance of our teleporter. Under d.c. excitation all photon pairs detected are not from the same radiative cascade, i.e. they are not correlated, and we formulate a probability that Bob detects D (conditional on Alice detecting H-V) taking uncorrelated emission at rate $\Gamma$ into account:

$$F_D(\tau_1, \tau_2) = \frac{P_{HVD}(\tau_1, \tau_2) + \Gamma \cdot \frac{1}{4}}{\left(P_{HVD}(\tau_1, \tau_2) + P_{HVA}(\tau_1, \tau_2)\right) + \Gamma \cdot \frac{1}{2}}$$

Similar expressions can be formulated for the degree of polarisation correlation [33] of the photon pair XX-X, which allows us to extract the parameters $\Gamma = 0.45 \text{ ns}^{-1}$ and $\gamma_x = 2.5 \text{ ns}^{-1}$ from independently measured polarisation correlation measurements.

Experimentally we measure third-order correlation functions $g^{(3)}_{HVD}(\tau_1, \tau_2)$ and $g^{(3)}_{HVA}(\tau_1, \tau_2)$, with Bob simultaneously recording the orthogonal polarisations D and A on his detectors, and we calculate the teleportation fidelity as

$$f_{D \to D}(\tau_1, \tau_2) = g^{(3)}_{HVD}(\tau_1, \tau_2) / \left(g^{(3)}_{HVA}(\tau_1, \tau_2) + g^{(3)}_{HVD}(\tau_1, \tau_2)\right)$$

To simulate the third-order correlation functions for the experimental setup at hand, we must take all cases that can lead to triple detections into account (similar to equation 1 above):

$$g^{(3)}_{HVD}(\tau_1, \tau_2) \propto \eta \alpha^2 \left(g^{(2)}_{VX}(\tau_2 - \tau_1) + g^{(2)}_{HX}(\tau_2)\right) F_D(\tau_1, \tau_2) + \eta^2 g^{(3)}_{HVD}(\tau_1, \tau_2)/2 + \alpha^4/4 \quad (2)$$

where $\alpha^2$ and $\eta$ are proportional to the laser and quantum dot XX intensities respectively, as measured on detectors D1 and D2. The term with $F_D(\tau_1, \tau_2)$ corresponds to the wanted case with one photon originating from the laser and one from the quantum dot. The second and third terms correspond to unwanted triples where all photons originate from the ELED or the laser, respectively. Again, full theoretical modelling of the quantum dot to calculate $g^{(3)}_{HVD}(\tau_1, \tau_2)$ is beyond the scope of this paper, and as in the case of $g^{(2)}_{HBT}(\tau)$ in equation 1 we use empirical fits that agree well with experimentally measured correlations.

After convolving with the experimentally determined detector time response functions, we find that the modelled third-order correlations have good qualitative agreement with measurements, and allow us to make the predictions of suitable laser to quantum dot intensity ratios and requirements on source detuning presented in Fig. 2b, as well as the simulations presented in Fig. 3.

**Acknowledgements**

The authors acknowledge partial financial support through the European Union Initial Training Network Spin Effects for Quantum Optoelectronics (SPIN-OPTRONICS) and the Seventh Framework Programme Future and Emerging Technologies Collaborative Project Quantum Interfaces, Sensors and Communication Based on Entanglement (Q-ESSENCE), the United Kingdom Engineering and Physical Sciences Research Council and the Cambridge Overseas Trust. The authors also thank C. Salter for support in device fabrication, and M. B. Ward for technical support.

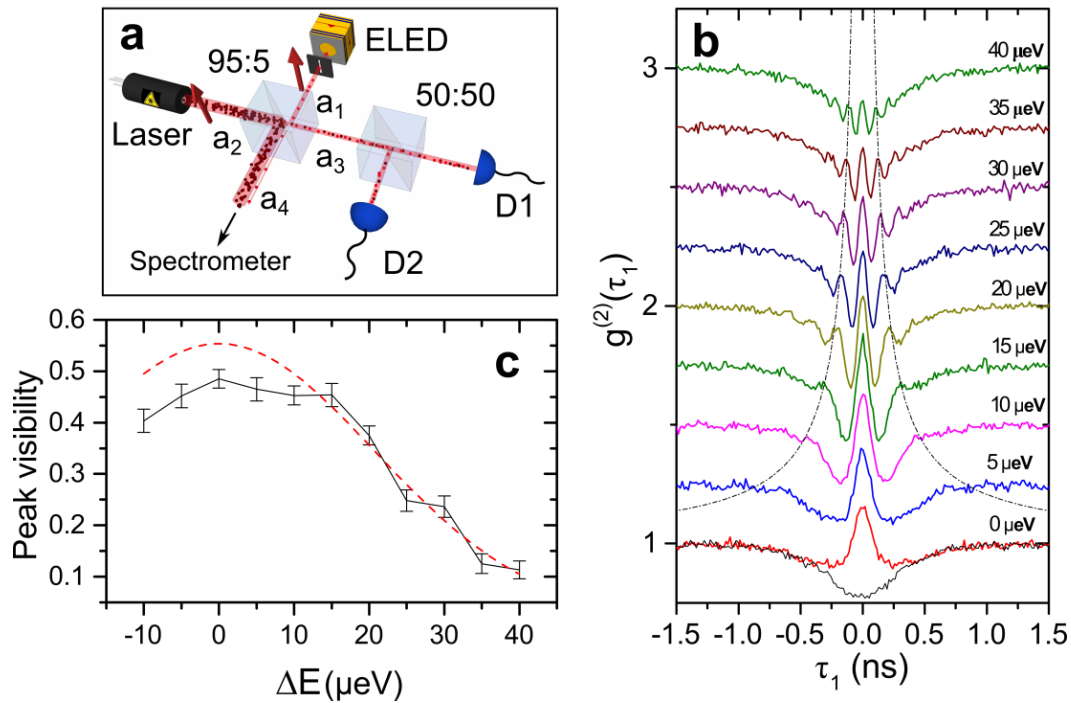

**Fig 1| Two-photon interference on an unbalanced beamsplitter.** (a) Schematic drawing of setup implemented in fibre optics. Modes $a_1$-$a_3$ couple with 95% efficiency, $a_2$-$a_3$ with 5%. A spectrometer in mode $a_4$ is used for monitoring purposes and controlling the source detunings. (b) Second-order correlation functions as a function of laser-biexciton photon detuning, exhibiting interference and quantum beats. Correlations with $\Delta E > 0$ have been offset for clarity. For $\Delta E = 0$ both interfering (red) and non-interfering (black) correlations are shown. Grey dashed line shows the predicted position of one quantum beat period. (c) Post-selected two-photon interference visibility as a function of detuning. Red dashed curve shows simulated visibility.

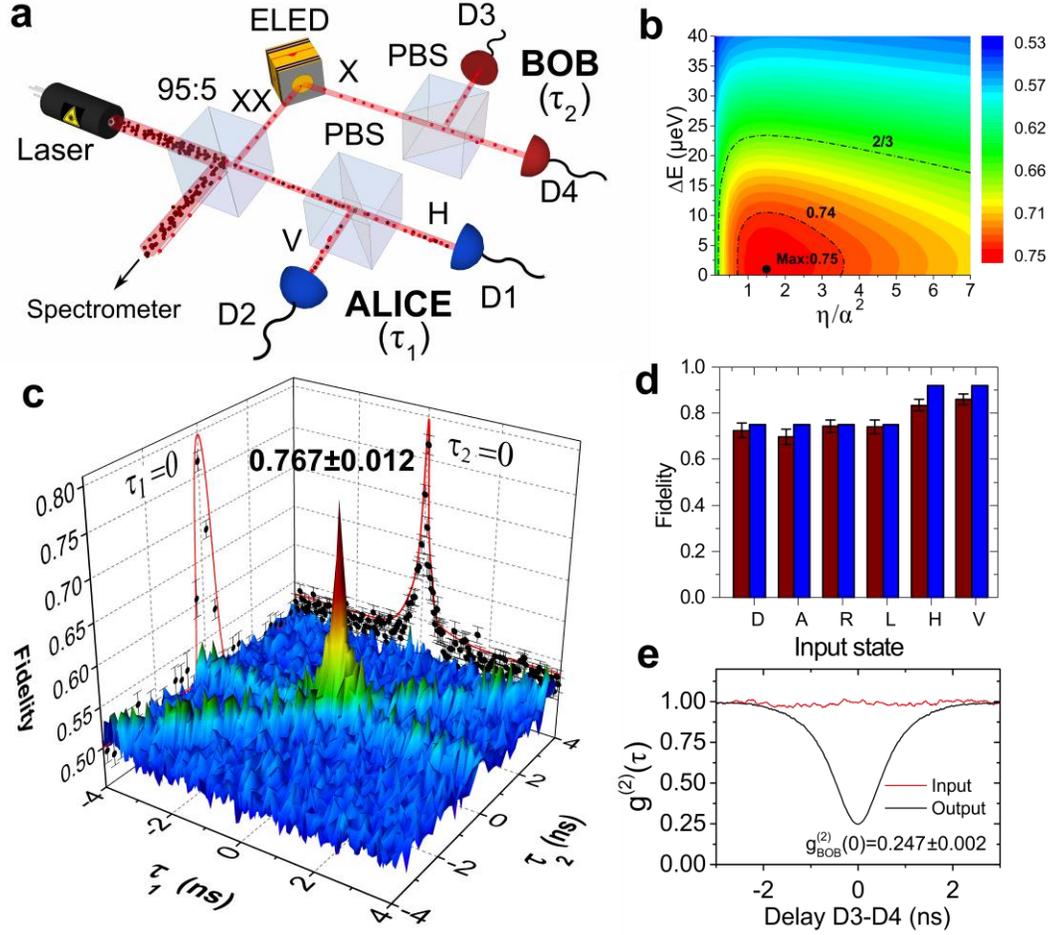

**Fig.2| Quantum teleportation of photons from a laser using an ELED**. (a) Schematic of experimental setup. (b) Influence of quantum dot to laser intensity ratio $\eta/\alpha^2$ and detuning $\Delta E$ on output fidelity for a superposition input state. (c) Surface plot showing average fidelity over six input states. Projections on sidewalls show cuts through experimentally measured 2D fidelity map for $\tau_1 = 0$ and $\tau_2 = 0$ (black dots), along with simulated fidelity (red curve). (d) Experimental (red) and simulated (blue) output fidelity for the six input states. (e) Photon statistics of the input (red) and output beams (black) of the teleporter.

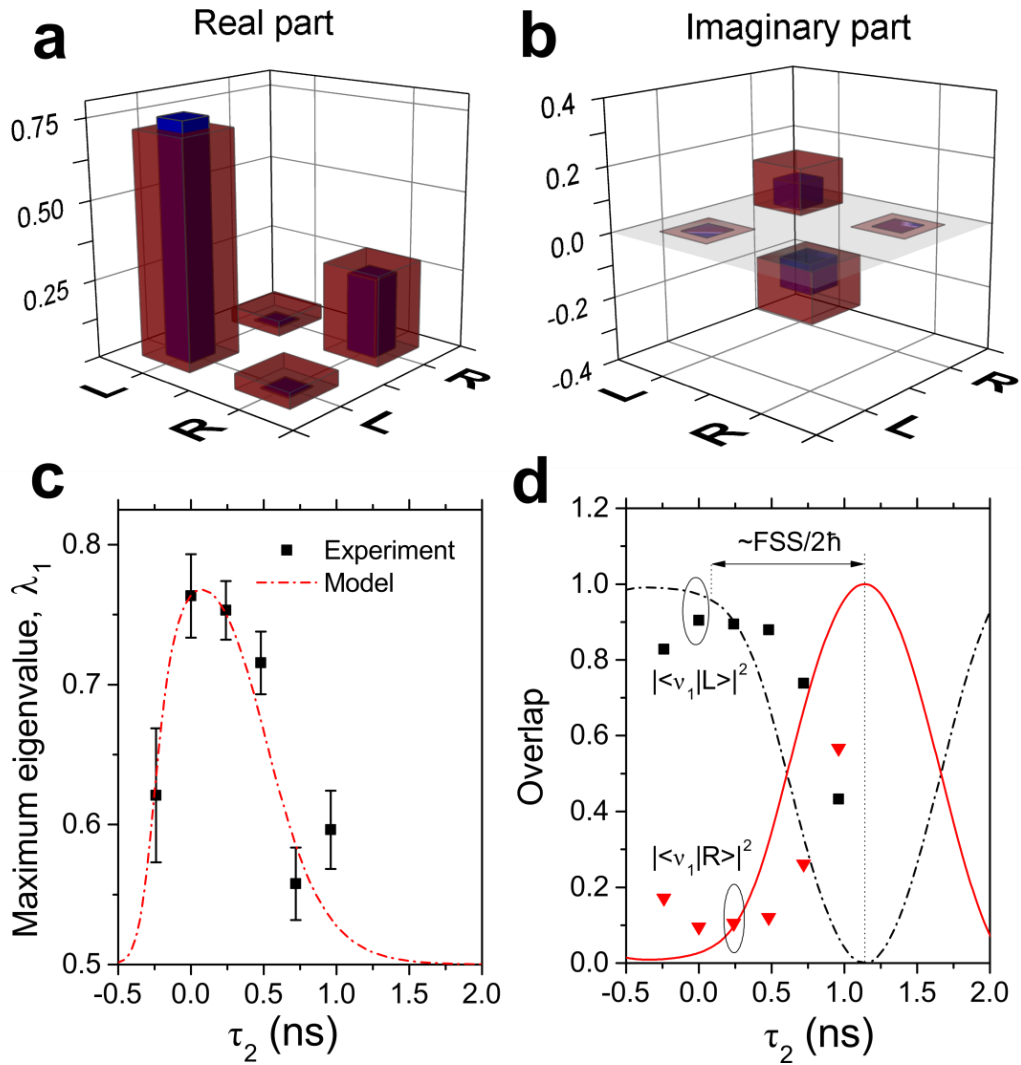

**Fig.3 | Evolution of teleported states.** Density matrix of photon received by Bob when teleporting R-polarised laser photons, showing real (a) and imaginary (b) parts respectively. Experimental data thick red bars, simulated thin blue bars. (c) Maximum eigenvalue of the density matrix as a function of Bob's detection time (d) Overlap between the optimal output state $|v_1\rangle$ with the desired output state L (black squares) and orthogonal state R (red triangles) respectively, showing time evolution with period determined by the exciton fine-structure splitting FSS~2 μeV.

13